\documentclass[aps,pra,twocolumn,showpacs,floatfix]{revtex4} 
\usepackage{amsmath,amssymb,isolatin1,graphicx,times}

\newcommand{\be}{\begin{equation}}
\newcommand{\ee}{\end{equation}}
\newcommand{\bea}{\begin{eqnarray}}
\newcommand{\eea}{\end{eqnarray}}
\newcommand{\bw}{\begin{widetext}}
\newcommand{\ew}{\end{widetext}}

\newcommand{\kommentar}[1]{}

\begin{document}
 
\title{Continuous time quantum walks in phase space}
\author{Oliver M{\"u}lken}
\author{Alexander Blumen}
\affiliation{
Theoretische Polymerphysik, Universit\"at Freiburg,
Hermann-Herder-Straße 3, D-79104 Freiburg, Germany}

\date{\today} 
\begin{abstract}
We formulate continuous time quantum walks (CTQW) in a discrete quantum
mechanical phase space. We define and calculate the Wigner function (WF)
and its marginal distributions for CTQWs on circles of arbitrary length
$N$. The WF of the CTQW shows characteristic features in phase space.
Revivals of the probability distributions found for continuous and for
discrete quantum carpets do manifest themselves as characteristic patterns
in phase space. 
\end{abstract}
\maketitle

\section{Introduction}

The study of quantum mechanical transport phenomena on discrete structures
is of eminent interest.  One particular aspect is the crossover from
quantum mechanical to classical behavior.  Since classical physics is
described in phase space and quantum mechanics in Hilbert space a unified
picture is desired. This is provided, for instance, by the so-called
Wigner function (WF) \cite{Wigner1932, Hillery1984}, which has remarkable
properties: It transforms the wave function of a quantum mechanical
particle into a function living in a position-momentum space similar to
the classical phase space. The WF is a real valued function
and in this respect compares well with the classical probability density
in phase space.  However, it is not always positive.

The concept of phase space functions has been widely used in Quantum
Optics \cite{Schleich,Mandel-Wolf} but also for describing electronic
transport, see e.g.\ \cite{kluksdahl1989,buot1993,bordone1999}. Here, the formal
similarity to the classical Boltzmann distribution has been exploited.
However, the phase spaces considered there are mostly continuous and
infinite. This results in a simpler mathematical tractability of WFs
than for discrete and for finite systems. Nevertheless, one can
also define WFs for discrete systems
\cite{Wootters1987,Cohendet1988,Leonhardt1995,Takami2001,Miquel2002}. 

Describing the transport by coined or continuous time quantum walks has
been very successful over the last few years, for an overview see
\cite{kempe2003}.  The applicability of quantum walks reaches far beyond
quantum computation; for instance in quantum optics also the (discrete)
Talbot effect can be described by continuous time quantum walks
\cite{iwanow2005,mb2005b}.
An additional, but more abstract approach to transport processes is
through the so-called quantum multibaker maps
\cite{wojcik2002,wojcik2003}. Such maps were shown to exhibit (as a
function of time) a quantum-classical crossover, where the crossover time
is given by the inverse of Planck's constant.

In the following we define the WF for a continuous time
quantum walk on a one-dimensional discrete network of arbitrary length $N$
with periodic boundary conditions (PBC). We further show that the marginal
distributions are correctly reproduced. Additionally, we define a long
time average of the WF, which we compare to the classical
limiting phase space distribution. Finally, we show how (partial) revivals
of the probability distribution manifest themselves in phase space.

\section{Continuous time quantum walks}

The quantum mechanical extension of a continuous time random walk (CTRW)
on a network (graph) of connected nodes
is called a continuous time quantum walk (CTQW). It is obtained by
identifying the Hamiltonian of the system with the (classical) transfer
matrix, ${\bf H} = - {\bf T}$, see e.g.\ \cite{farhi1998,mb2005a} (we will set
$\hbar \equiv 1$ in the following). The transfer matrix of the walk, ${\bf
T} = (T_{ij})$, is related to the adjacency matrix ${\bf A}$ of the graph
by ${\bf T} = - \gamma {\bf A}$, where for simplicity we assume the
transmission rate $\gamma$ of all bonds to be equal.  The matrix ${\bf A}$
has as non-diagonal elements $A_{ij}$ the values $-1$ if nodes $i$ and $j$
of the graph are connected by a bond and $0$ otherwise.  The diagonal
elements $A_{ii}$ of ${\bf A}$ equal the number of bonds $f_i$ which exit
from node $i$.

The basis vectors $|j\rangle$ associated with the nodes $j$ span the whole
accessible Hilbert space to be considered here. The time evolution of a
state $| j \rangle$ starting at time $t_0$ is given by $| j;t \rangle =
{\bf U}(t,t_0) | j \rangle$, where ${\bf U}(t,t_0) = \exp[-i {\bf H}
(t-t_0)]$ is the quantum mechanical time evolution operator.

\section{Wigner functions}

\subsection{Definition}

The WF is a quasi-probability (in the sense that it can become
negative) in the quantum mechanical phase space.  For a $d$ dimensional
system, the (quantum mechanical) phase space is $2d$ dimensional.  If the
phase space is spanned by the continuous variables $X$ and $K$, the WF is
given by \cite{Wigner1932,Hillery1984}
\be
W(X,K;t) = \frac{1}{\pi}\int \ dY \ e^{iKY} \ \langle X - Y/2 | \hat
\rho(t) | X + Y/2
\rangle ,
\label{wigner}
\ee
where $\hat \rho(t)$ is the density operator and thus for a pure state,
$\hat \rho (t) = | \psi(t) \rangle \langle \psi(t) |$. Here, $\psi(X;t) =
\langle X | \psi(t) \rangle$ is the wave function of the particle.
Integrating $W(X,K;t)$ along lines in phase space gives marginal
distributions, e.g., when integrating along the $K$-axis one has,
\cite{Wigner1932,Hillery1984},
\be
\int d K \ W(X,K;t) = |\psi(X;t)|^2.
\label{marg_x}
\ee

In the case of a discrete system, given, for instance, by $N$ discrete
positions on a network, which we choose to enumerate as $0,1,\dots,N-1$,
the functions $\psi(x)$ are only defined for integer values of
$x=0,1,\dots,N-1$, and the form of Eq.(\ref{wigner}) has to be changed
from an integral to a sum.  There have been several attempts in doing so,
see, for instance, \cite{Wootters1987,Cohendet1988}. However, the
definition of the discrete WFs might depend on whether the length $N$ of
the system is even or odd \cite{Leonhardt1995,Leonhardt1996}. 

In the following we proceed somewhat differently than previous works and
focus on discrete systems with PBC. For a one-dimensional system of length
$N$ (exemplified by a ring) this implies $\psi(x)\equiv \psi(x\pm rN)$ for
all $r\in\mathbb{N}$. It follows that each and every one of the products
$\psi^*(x-y';t) \psi(x+y';t)$ is identical to (at least) one of the $N$
forms $\psi^*(x-y;t) \psi(x+y;t)$, where $y=0,1,\dots,N-1$.

Now the WF has the form of a Fourier transform; a unique transformation of
these $N$ products requires $N$ different $k$-values. These $k$-values may
evidently be chosen as $k=2\pi\kappa/N$, again having
$\kappa=0,1,\dots,N-1$. We are thus led to propose for integer $x$ and $y$
the following discrete WF
\be
W(x,k;t) = \frac{1}{N} \sum_{y=0}^{N-1} e^{iky} \ \psi^*(x-y;t)
\psi(x+y;t).
\label{wigner_discrete}
\ee

As a side remark we note that the rhs of Eq.(\ref{wigner_discrete}) stays
invariant when replacing $N$ by $mN$ (with $m$ integer and $m\neq 0$).
Furthermore, also the choice of the summation interval $0,1,\dots,N-1$ is
arbitrary; any $N$ consecutive $y$-values will do. Given that the nature
of the Hilbert space underlying Eq.(\ref{wigner}) differs from that of
Eq.(\ref{wigner_discrete}), there is no simple relation connecting the two
equations. Here, one may recall the problem of the normalization of plane
waves in bounded (compact) and in infinite spaces. In the following we will
use Eq.(\ref{wigner_discrete}) in order to compute the WFs.

\subsection{WFs and the Bloch ansatz }\label{wigner_bloch_odd}

In the following we consider only CTQWs with nearest neighbor steps.  The
Hamilton operator for such a CTQW on a finite one-dimensional network with
PBC takes on a very simple form
\be
{\bf H} |j\rangle = 2 |j\rangle - |j-1\rangle - |j+1\rangle,
\label{hamil} 
\ee
where we have taken the transmission rate to be $\gamma\equiv 1$. The states
$|j\rangle$, where $j=0,1,\dots,N-1$, are the basis states associated with
the nodes $j$. Since here the CTQW describes a particle moving in a
periodic potential we can employ the Bloch ansatz \cite{Ziman,mb2005b}.
The time independent Schr\"odinger equation ${\bf H} |\Phi_\theta\rangle =
E_\theta |\Phi_\theta\rangle$ has eigenstates $|\Phi_\theta\rangle$ which
are Bloch states. The PBC require that $\Phi_\theta(N) = \Phi_\theta(0)$,
where $\Phi_\theta(x) = \langle x | \Phi_\theta \rangle$. As before, this
restricts the $\theta$-values to $\theta=2\pi n/N$, where $n=0,1,\dots,N-1$. The
Bloch state $|\Phi_\theta\rangle$ can be expressed as a linear combination
of the states $|j\rangle$,
\be
|\Phi_\theta\rangle = \frac{1}{\sqrt N}\sum_{j=0}^{N-1} e^{i \theta
j} |j\rangle.
\ee
In turn, the state $|j\rangle$ can be expressed by the Bloch states,
i.e.\ by a Wannier function \cite{Ziman},
\be
|j\rangle = \frac{1}{\sqrt N}\sum_{\theta} e^{-i \theta
j} |\Phi_\theta\rangle, 
\label{blochef}
\ee
where the sum runs over the $N$ values for $\theta$.  From this it follows
readily that the eigenvalues to Eq.(\ref{hamil}) are given by $E_\theta =
2 - 2 \cos\theta$.  The time evolved state then follows as
\be
|j;t\rangle \equiv e^{-i{\bf H}t} | j\rangle = \frac{1}{\sqrt N}
\sum_\theta e^{-i\theta j} e^{-i E_\theta t} | \Phi_\theta \rangle,
\label{timeevol_j}
\ee
Using the position basis means introducing $\psi_j(x;t) \equiv \langle x |
j;t \rangle$. Evidently, at $t=0$ one has  $\psi_j(x;0) = \langle x | j
\rangle = \delta_{x,j}$. Furthermore, with Eq.(\ref{timeevol_j}) we have
\bea
&& \psi^*_j(x';t)\psi_j(x;t) \equiv \langle j;t |
x'\rangle\langle x | j;t\rangle 
\nonumber \\
&& = \frac{1}{N} \sum_{\theta,\theta'}
e^{-i(\theta-\theta')j} e^{i (E_{\theta'} - E_\theta) t}
\Phi^*_{\theta'}(x') \Phi_\theta(x)
\nonumber \\
&& = \frac{1}{N^2} \sum_{\theta,\theta'}
e^{-i(\theta-\theta')j} e^{i (E_{\theta'} - E_\theta) t} 
\sum_{l,l'} e^{il\theta} e^{-il'\theta'}
\ \delta_{x',l'}\delta_{x,l}
\nonumber \\
&& = \frac{1}{N^2} \sum_{\theta,\theta'}
e^{-i(\theta-\theta')j} e^{i (E_{\theta'} - E_\theta) t} e^{i\theta x}
e^{-i\theta' x'}.
\label{wigner_kernel}
\eea

Now, following Eq.(\ref{wigner_discrete}), the WF is given by
\bea
W_j(x,k;t) 
&=& 
\frac{1}{N^3} \sum_{\theta,\theta'} e^{i(\theta-\theta')(x-j)} \
e^{i(E_{\theta'} - E_\theta)t} 
\nonumber \\
&&\times 
\sum_y e^{i(\theta+\theta'+k)y} 
.
\label{wigner_bloch_0}
\eea
The sum over $y$ can be written as $N\Delta_{\theta+\theta'+k}$ where, as
an extension of the usual Kronecker symbol, we introduce $\Delta_n$, such
that $\Delta_n=1$ for $n\equiv0$ (mod $N$) and $\Delta_n=0$ else.
Therefore, the WF for a CTQW on a one-dimensional network of $N$ nodes
with PBC reads
\bea
&& W_j(x,\kappa;t) = \frac{1}{N^2} \sum_{n=0}^{N-1}
\exp\big[
-i2\pi(2n+\kappa)(x-j)/N
\big]
\nonumber \\
&& \times 
\exp\big\{
-i2t[\cos(2\pi (\kappa + n)/N) - \cos(2\pi n/N)]
\big\}
,
\label{wigner_bloch_3}
\eea
where we have used that $\theta=2\pi n/N$ and $k=2\pi \kappa/N$.
Furthermore, one may note from Eq.(\ref{wigner_discrete}) that
\bea
\sum_\kappa
W_j(x,\kappa;t) &=& \frac{1}{N} \sum_\kappa \sum_y e^{i2\pi\kappa y/N}
\nonumber \\
&& \times 
\psi^*(x-y;t) \psi(x+y;t)
\nonumber \\
&=& 
|\psi(x;t)|^2.
\label{wigner_disc_marg}
\eea

Note that the WF in Eq.(\ref{wigner_bloch_3}) does not change with time for $\kappa=0$, i.e.\ we have
$W_j(x,0;t)=W_j(x,0;0)= {\cal W}_j(x,0)$, where ${\cal W}_j(x,\kappa)$ is
the long time average of the WF given in Sec.\ref{longtimeavg}.

\section{WFs and CTQWs}

In the following we consider WFs for CTQWs on circles of different length.
For $t=0$ we start the CTQWs from one node. In particular, we take $N=100$
(as an example of an even value) and $N=101$ (as odd value); as starting
site we choose $j=50$.
 
\subsection{CTQWs in phase space}

\begin{figure}[ht]
\centerline{\includegraphics[clip=,width=0.9\columnwidth]{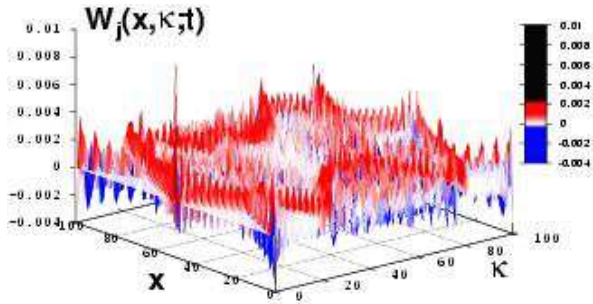}}
\caption{(Color online) 3D visualization of the WFs of a CTQW
on a cycle of length $N=101$ at $t=40$. The initial node is at $j=50$.}
\label{wigner_bloch_101_3d}
\end{figure}

\begin{figure}[ht]
\centerline{\includegraphics[clip=,width=0.9\columnwidth]{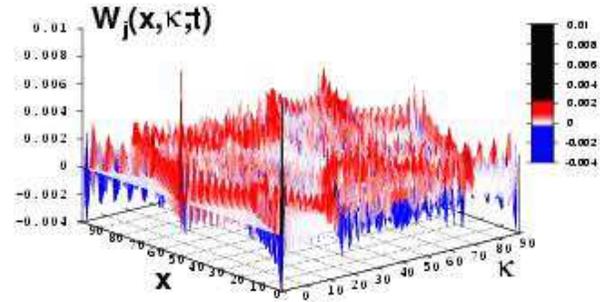}}
\caption{(Color online) Same as Fig.\ref{wigner_bloch_101_3d}, for $N=100$
at $t=40$ with $j=50$.}
\label{wigner_bloch_100_3d}
\end{figure}

In general, the WFs have a very complex structure.  Figures
\ref{wigner_bloch_101_3d} and \ref{wigner_bloch_100_3d} display the WFs
for CTQWs over cycles with $N=101$ and $N=100$, respectively, at $t=40$.
The peaks at $(x,\kappa)=(j,0)$ for $N=101$ and for $N=100$ are clearly
noticeable (black peaks in Figs.\ref{wigner_bloch_101_3d} and
\ref{wigner_bloch_100_3d}); however, the remaining structure is quite
complex.

\begin{figure}[ht]
\centerline{\includegraphics[clip=,width=0.94\columnwidth]{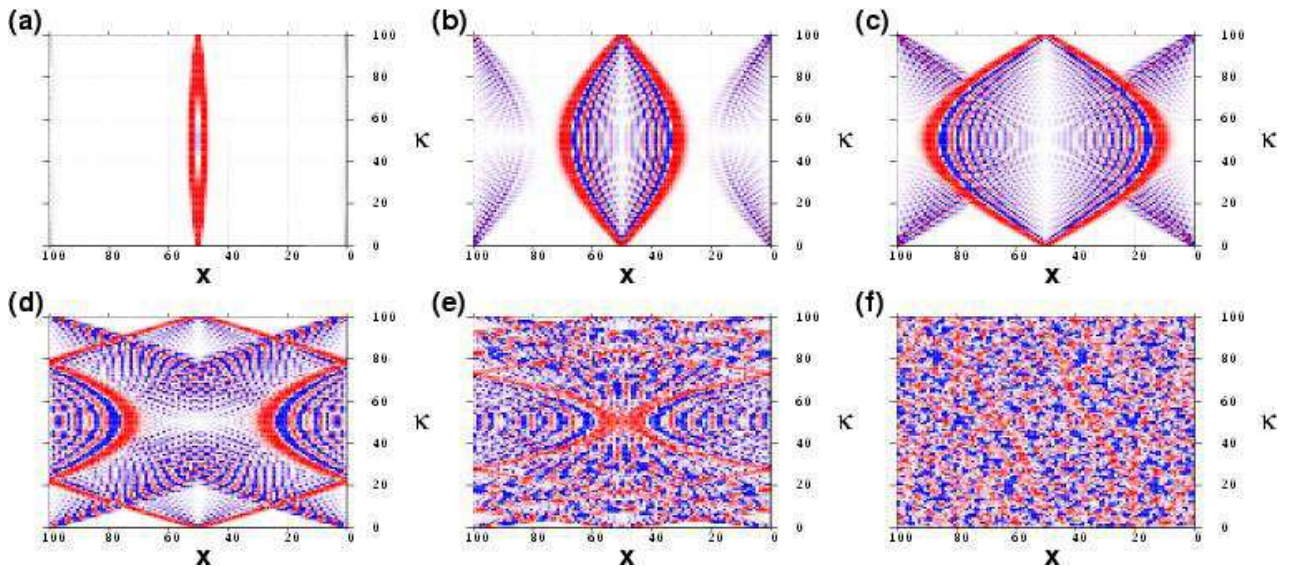}}
\caption{(Color online) WFs of a CTQW on a cycle of length
$N=101$ at times $t=1,10,20,40$ [(a)-(d)] as well as $t=100,500$
[(e),(f)]. The initial node is at $j=50$.  Red regions denote positive
values of $W_j(x,\kappa;t)$, blue regions negative values and white
regions values close to $0$.
}
\label{wigner_bloch_101_time}
\end{figure}

Figure \ref{wigner_bloch_101_time} shows a contour plot of the WF of a
CTQW on a cycle of length $N=101$ at different times. Note that at $t=0$
the WF is localized on the strip at the initial point $j$. At $t=1$, the
WF is still mostly localized about $j$. As time increases other sites get
populated. On short time scales, the WF develops a very regular structure
in phase space, with ``wavefronts'' originating from the initial point
$j$.  Additionally, we also note ``wavefronts'' starting from the region
opposite to the initial point, which are much weaker in amplitude.  As
time progresses, these two types of waves start ``interfering'' with each
other.

\begin{figure}[ht]
\centerline{\includegraphics[clip=,width=0.94\columnwidth]{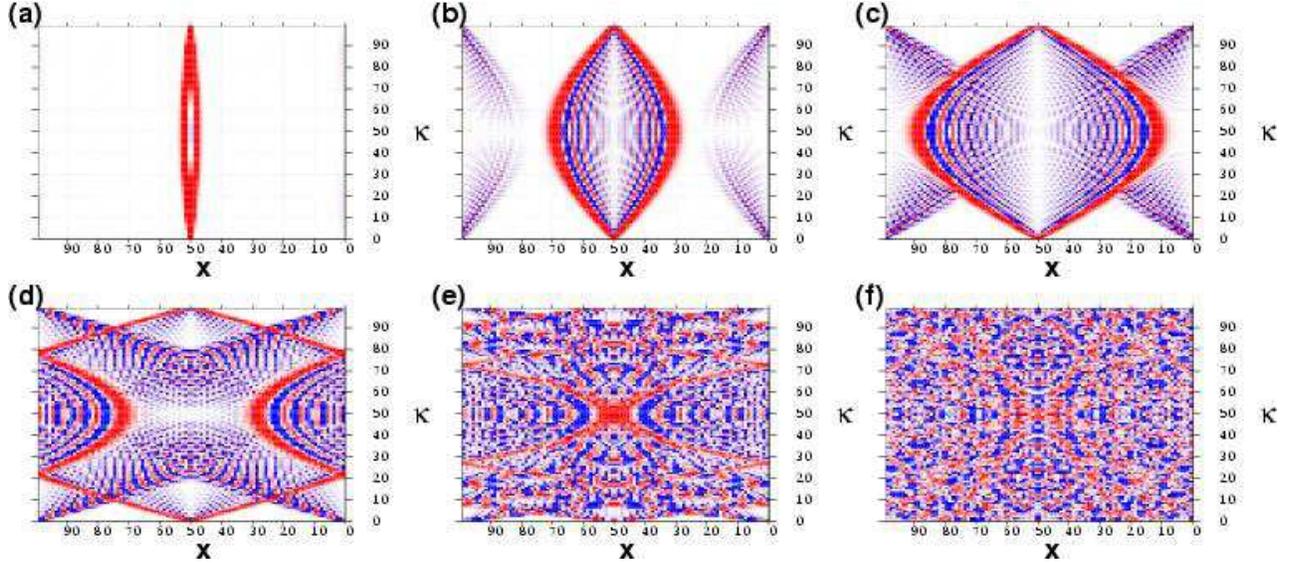}}
\caption{(Color online) Same as Fig.\ref{wigner_bloch_101_time}, for
$N=100$ and $j=50$.}
\label{wigner_bloch_100_time}
\end{figure}

In Fig.\ref{wigner_bloch_100_time} we plot the WF of a CTQW for $N=100$.
The structure of the WF is quite similar to the odd numbered network.
Nonetheless, there are differences at larger times, visible by comparing
Figs.\ref{wigner_bloch_101_time}(e) and \ref{wigner_bloch_101_time}(f) to
Figs.\ref{wigner_bloch_100_time}(e) and \ref{wigner_bloch_100_time}(f).

\bw
~
\ew

\bw
~
\ew

However, although at long times the interference effects are quite
intricate, typical patterns are still visible.  At $t=500$, we find less
regularities in phase space for $N=101$ than for $N=100$, reflecting the
higher symmetry of CTQWs on even numbered networks. We note moreover, that
the phase space pictures give at any time a much richer picture of the
underlying dynamics than the transition probabilities $|\psi_j(x;t)|^2$
alone. While the transition probabilities appear to be quite irregular
\cite{mb2005b}, the WFs display regular patterns. As in the case of a
particle in the box \cite{friesch2000}, also here we might view the
ensuing structure, the ``quantum carpets'', to be woven by the discrete
WFs. 

We finally note that in an infinite system the WFs have a much simpler
structure, because there we do not face the problem of distinct wave
fronts running into opposite directions and interfering with each other.

\subsection{Marginal distributions}

In general, the marginal distribution $\sum_\kappa W_j(x,\kappa;t)$ is
obtained from Eq.(\ref{wigner_bloch_3}) as
\bea
\sum_\kappa
W_j(x,\kappa;t)
&=& \frac{1}{N} \sum_{n=0}^{N-1}\exp[-i2\pi n(x-j)/N] 
\nonumber \\
&&\times \exp[i2t\cos(2\pi
n/N)]
\nonumber \\
&\times&
\frac{1}{N} \sum_{\kappa=0}^{N-1}\exp[-i2\pi(\kappa+n)(x-j)/N]
\nonumber \\
&&\times \exp[-i2t\cos(2\pi (\kappa
+ n)/N)].
\eea
Since the system is periodic, the arguments $(\kappa+n)$ in the
exponentials can be substituted by $(\kappa+n)\equiv~N-\nu$, where
$\nu=0,1,\dots,N-1$. This yields
\be
\sum_\kappa
W_j(x,\kappa;t)
=\left| \frac{1}{N} \sum_{\nu=0}^{N-1} e^{i2\pi \nu(x-j)/N}
e^{-i2t\cos(2\pi \nu/N)}
\right|^2, 
\label{marg_x_o}
\ee
which is exactly what also follows, see Eq.(\ref{wigner_disc_marg}), from
calculating $|\psi(x;t)|^2$ directly
from the Bloch ansatz, see Eq.(\ref{wigner_kernel}). The same relation was
also highlighted in Ref.\cite{mb2005b}.

Furthermore, we obtain the marginal distribution $\sum_\kappa
W_j(x,\kappa;t)$ for the infinite system by taking $N$ large (but fixed)
in Eq.(\ref{marg_x_o}), and setting $\theta=2\pi \nu/N$. We are then led to 
\bea
\sum_\kappa W_j(x,\kappa;t) &=& \left|\frac{1}{2\pi} \int\limits_0^{2\pi}
d\theta \ e^{i\theta(x-j)} e^{-i2t\cos\theta} \right|^2 
\nonumber \\
& = &
\left[J_{x-j}(2t)\right]^2,
\label{marg_x_inf}
\eea
where $J_\nu(\tau)$ is the Bessel function of the first kind, see
Eq.(9.1.21) of Ref.\cite{abramowitz}.

For the marginal distribution
$\sum_x W_j(x,\kappa;t)$ we also get from
Eq.(\ref{wigner_bloch_3})
\bea
&&\sum_{x} W_j(x,\kappa;t) 
= \frac{1}{N^2} \sum_{n=0}^{N-1} N\Delta_{2n+\kappa}
\ e^{i2\pi(2n+\kappa)j/N} 
\nonumber \\
&& \times \exp\big\{
-i2t[\cos(2\pi (\kappa + n)/N) - \cos(2\pi n/N)]
\big\}
\nonumber \\
&&= \begin{cases} 1/N & N \ \mbox{odd and all} \ \kappa \\
2/N & N \ \mbox{even and} \ \kappa \ \mbox{even} \\
0 &  N \ \mbox{even and} \ \kappa \ \mbox{odd},
\end{cases}
\label{marg_k_o}
\eea
which is independent of $t$. Eq.(\ref{marg_k_o}) can be confirmed directly
by taking the Fourier transform of $|\psi(x;t)|^2$.  The whole phase space
volume is normalized to one, as can be seen by summing Eq.(\ref{marg_k_o})
over all $\kappa$, with $\kappa=0,1,\dots,N-1$.

\subsection{Long time averages}\label{longtimeavg}

Similar to the classical case, the CTQW will visit any point in phase
space. However, due to the evolution with the unitary ${\bf H}$, there is
no definite limiting distribution. For the CTQW one may view the long time
average of the transition probability $|\psi_j(x;t)|^2$ as a limiting
distribution \cite{aharonov2001}, i.e.
\be
\chi_{xj} \equiv \lim_{T\to\infty}\frac{1}{T} \int\limits_0^T dt \
|\psi_j(x;t)|^2.
\ee 
Employing this idea, we define the limiting WF as
\be
{\cal W}_j(x,\kappa)
\equiv \lim_{T\to\infty}\frac{1}{T} \int\limits_0^T dt \
W_j(x,\kappa;t).
\ee
Upon integration of Eq.~(\ref{wigner_bloch_3}) over $t$ the only term
which remains is the one for which $\cos[2\pi(\kappa+n)/N] = \cos[2\pi
n/N]$. 

Some care is in order here, since for the long time average there are
differences between even and odd $N$, also occurring in the limiting
distribution $\chi_{xj}$ \cite{mb2005b}.  In the case of an odd $N$
(superscript $^o$) we get 
\be
{\cal W}^o_j(x,\kappa)
= \begin{cases} 
1/N^2 & \kappa\neq 0 \ \mbox{and any} \ x \\ 
1/N & \kappa=0 \ \mbox{and} \
x=j \\ 
0 &
\mbox{else}. \end{cases}
\ee
Note that for $k\neq0$ we require $2n=N-\kappa$. 

For an even $N$ (superscript $^e$), the limiting WF for $\kappa=0$ has
contributions only for $x=j$ {\sl and} $x=j+ N/2$. Each of the two
contributions has weight $1/2$. Thus the limiting WF reads
\be
{\cal W}^e_j(x,\kappa)
= \begin{cases} 
1/N^2 & \kappa\neq0 \ \mbox{and any} \ x \\
1/2N & \kappa=0 \
\mbox{and} \ x=j,j+ N/2 \\ 
0 & \mbox{else}. \end{cases}
\ee
Note that also here we require that $2n=(N-k)$.

The long time average of the WF is always positive. Note that for
classical CTRWs the limiting phase space distribution is uniform and
equals $1/N^2$ both for odd {\sl and} for even $N$.  Thus, what
distinguishes the limiting WF from the classical limiting distribution are
a few exceptional points in the phase space.

The long time average of the marginal distribution $|\psi_j(x;t)|^2$ is
also recovered from the limiting WF as $\sum_\kappa \ 
{\cal W}_j(x,\kappa)
= \chi_{xj}$, thus for odd $N$
\be
\chi^o_{xj} = \begin{cases} (2N-1)/N^2 & \mbox{for} \ x=j \\ (N-1)/N^2 &
\mbox{else}\end{cases}
\ee
and for even $N$
\be
\chi^e_{xj} = \begin{cases} (2N-2)/N^2 & \mbox{for} \ x=j,j+ N/2 \\
(N-2)/N^2 & \mbox{else}\end{cases},
\ee
which confirms our previous calculations \cite{mb2005b}.

\subsection{Revivals in phase space}

CTQWs on large networks show only partial revivals \cite{mb2005b}. These
should occur at times $t_r\gtrapprox N^2/2\pi$. However, for small
networks of sizes $N=1,2,3,4,6$ also complete revivals are possible, as
was recently seen in waveguide arrays \cite{iwanow2005}. The WFs are
easily obtained from Eq.(\ref{wigner_bloch_3}). The marginal distributions
$\sum_\kappa W_j(x,\kappa;t)$ are, of course, correctly recovered. For
instance, for $N=6$, the first complete revival occurs for $t=2\pi$. This
is readily seen, since for the WF at $x=j$ we have $W_j(x=j,\kappa;t=2\pi)
= 1/6$; thus the marginal distribution is $\sum_\kappa
W_j(x=j,\kappa;t=2\pi) = 1$, which also means, using
Eq.(\ref{wigner_disc_marg}), that $|\psi_j(x=j;t=2\pi)|^2=1$.

\begin{figure}[ht]
\centerline{\includegraphics[clip=,width=\columnwidth]{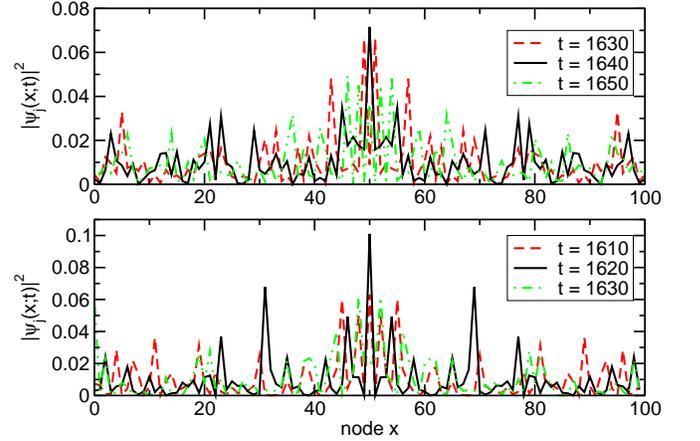}}
\caption{Transition probability $|\psi_j(x;t)|^2$ as function of $x$ of
CTQWs on cycles of length (a) $N=101$ at times $t=1620,1640,1650$ and (b)
$N=100$ at times $t=1610,1620,1630$. The initial node is $j=50$ in both
cases.}
\label{wigner_x_t}
\end{figure}

\begin{figure}[ht]
\centerline{\includegraphics[clip=,width=0.9\columnwidth]{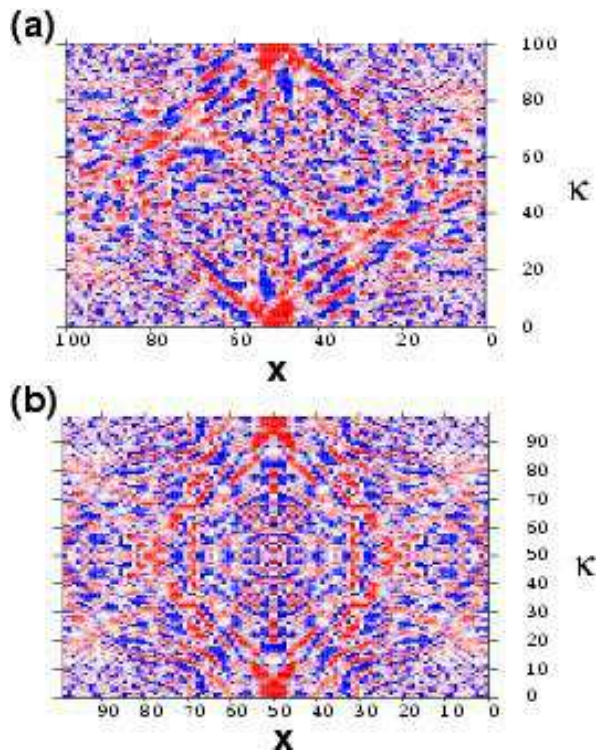}}
\caption{(Color online) WFs of a CTQW on a cycle of length
(a) $N=101$ at time $t=1640$ and (b) $N=100$ at time $t=1620$. The initial
node is $j=50$ in both cases.  Red regions denote positive values, blue
regions negative values and white regions values close to $0$ of
$W_j(x,\kappa;t)$.}
\label{wigner_bloch_101_revival}
\end{figure}

Figure \ref{wigner_x_t} shows the transition probability $|\psi_j(x;t)|^2$
for $N=101$ and $N=100$ at three different times $t$ close to the
corresponding $t_r$. For $N=101$ ($N=100$) at $t=1640$ ($t=1620$) the CTQW
shows the closest revival to the initial distribution $|\psi_j(x;0)|^2$ in
that time interval. The corresponding phase space pictures are given in
Fig.\ref{wigner_bloch_101_revival}. 

Although the phase space pictures in Fig.\ref{wigner_bloch_101_revival}
are vastly different from the ones at the initial time, there are strong
peaks in the WFs seen as red regions close to the initial point, $j=50$.
Moreover, the patterns in Figs.\ref{wigner_bloch_101_revival}(a) and (b)
are more regular than, say, the ones in
Figs.\ref{wigner_bloch_101_time}(f) and \ref{wigner_bloch_100_time}(f).
Especially along lines perpendicular to the initial node $j$ there are in
both cases areas with strong positive peaks in the WF, giving rise to
strong peaks in the marginal distribution $\sum_\kappa W_j(x,\kappa;t)$.
All this means that close to a partial revival the structure is by far
more regular than at an arbitrary moment in time.

\section{Conclusions}

In conclusion, we have defined and calculated the discrete WF of a CTQW on
a discrete cycle of arbitrary length $N$ with PBC. Different from previous
attempts, our definition is independent of whether $N$ is even or odd.
Integrating the WF along lines in phase space gives the correct marginal
distributions. The WF exhibits characteristic patterns.  Furthermore, we
also showed how partial revivals of the WF manifest themselves in phase
space. The patterns in phase space allow us to draw a much richer picture
of the CTQW than the marginal distributions can. 

\section*{Acknowledgments}

This work was supported by a grant from the Ministry of Science, Research
and the Arts of Baden-W\"urttemberg (AZ: 24-7532.23-11-11/1). Further
support from the Deutsche Forschungsgemeinschaft (DFG) and the Fonds der
Chemischen Industrie is gratefully acknowledged.

\end{document}